\documentclass[11pt,twoside]{article}
\usepackage{asp2010}

\resetcounters
\newcommand\teff{T_{\rm eff}}
\newcommand{\wig}[1]{\mathrel{\hbox{\hbox to 0pt{%
     \lower.6ex\hbox{$\sim$}\hss}\raise.4ex\hbox{$#1$}}}}
\begin{document}

\title{Towards an Understanding of the Atmospheres of Cool White Dwarfs}
\author{Piotr M. Kowalski
$^{1,2}$, Didier Saumon$^3$, Jay Holberg$^4$, Sandy Leggett$^5$}

\affil{$^1$ IEK-6 Institute of Energy and Climate Research, Forschungszentrum J\"{u}lich, 52425 J\"{u}lich, Germany}
\affil{$^2$ 
GFZ German Research Centre for Geosciences, Telegrafenberg, 14473 Potsdam, Germany}
\affil{$^3$ 
Los Alamos National Laboratory, Mail Stop F663, Los Alamos NM 87545, USA}
\affil{$^4$ Lunar and Planetary Laboratory, University of Arizona, Tucson, Arizona, 85721, USA}
\affil{$^5$ Gemini Observatory, 670 N. A\'ohoku Place, Hilo, HI 96720, USA}

\begin{abstract}
Cool white dwarfs with $T_{\rm eff}\wig< 6000\,$K are the remnants of the oldest stars that existed in our Galaxy.
Their atmospheres, when properly characterized, can provide valuable information on white dwarf evolution
and ultimately star formation through the history of the Milky Way. Understanding the atmospheres of these stars 
requires joined observational effort and reliable atmosphere modeling.
We discuss and analyze recent observations of the near-ultraviolet (UV) and near-infrared (IR) spectrum of several cool white dwarfs 
including DQ/DQp stars showing carbon in their spectra.
We present fits to the entire spectral energy distribution (SED) of selected cool stars, 
showing that the current pure-hydrogen atmosphere models are quite reliable, especially in the near-UV spectral region.
Recently, we also performed an analysis 
of the coolest known DQ/DQp stars investigating further the origin of the $\rm C_2$ Swan bands-like spectral features 
that characterize the DQp stars. We show that the carbon abundances derived for DQp stars fit the trend of carbon abundance with $T_{\rm eff}$ 
seen in normal cool DQ stars. This further supports the recent conclusion of \citet{K10} that
DQp stars are DQ stars with pressure distorted Swan bands. 
However, we encounter some difficulties in reproducing the IR part of the SED of stars
having a mixed He/H atmosphere. This indicates limitations in current  models of the opacity in dense He/H fluids.
\end{abstract}

\section{Introduction}
Being billions of years old, cool white dwarfs (WD) represent the last stage of the evolution of most stars.
When properly understood these stellar remnants may be used to investigate stellar evolution,
cosmochronometry \citep{FBB01,R06} and even to determine the bulk composition of exoplanetary material thought to be accreting on the surfaces of many
WD (e.g. \citet{F10}). In order to extract the valuable information contained in the spectra of cool WD, we have to properly understand
their atmospheres \citep{KWD10} where  more extreme physical conditions are encountered than in normal stars. In particular, He-dominated 
atmospheres have fluid-like densities (up to a few $\rm g/cm^3$, \citet{KK10}). Over the last decade many dense-medium effects were introduced
into the modeling, substantially increasing fidelity of the atmosphere models. This has resulted in significant improvements in our understanding 
of these stars (see \citet{KK10} and \citet{KWD10}). Here we briefly discuss the current state of understanding of cool white dwarfs and the results 
of our most recent studies.

\begin{figure}
\resizebox{\hsize}{!}{\rotatebox{270}{\includegraphics{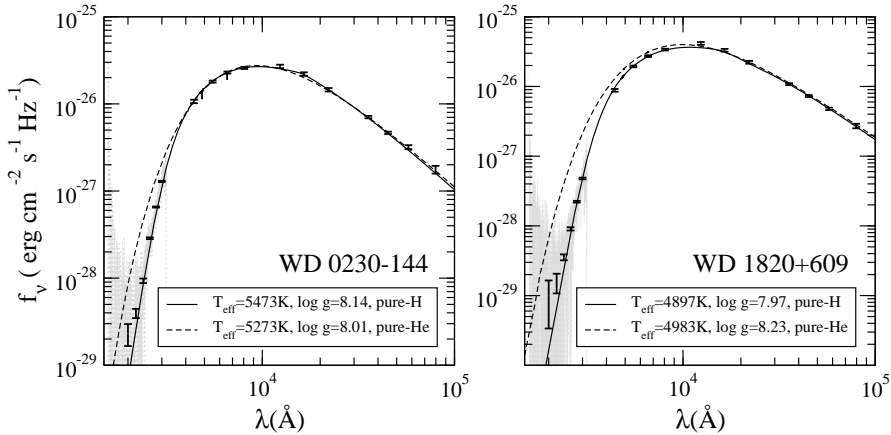}}}
  \caption{The measured SED of two cool white dwarfs. Solid and dashed lines represent the fits with  pure-H and pure-He atmosphere models,
           respectively.  Data shortward of 4000$\,\AA$ were taken with the STIS instrument on HST. \label{F1}}
\end{figure}

\section{Atmospheres of hydrogen and helium }

\subsection{Test of pure-hydrogen models}

With the inclusion of the red wing of the Ly-$\alpha$ line and improvements in the description of the physics of He-rich atmospheres,
we obtained a qualitatively different
set of atmosphere models whose application in the analysis of data has resulted in better determinations
of the surface composition of cool H/He WD \citep{KK10}. The inclusion of Ly-$\alpha$ absorption \citep{KS06} led to
a successful reproduction of the entire spectral energy distribution (SED) of many cool white dwarfs (from the near-UV to the IR, e.g. \citet{KK09,KS06}). However, until recently the spectra 
in UV region, where the Ly-$\alpha$ is the dominant opacity, had been measured for only two stars with hydrogen-rich atmospheres, which we successfully reproduced with our models 
\citep{KS06,DK12}. To further test the reliability of the Ly-$\alpha$ line profile model of \citet{KS06}
over a broader range of conditions, we obtained near-UV spectra of eight cool white dwarfs with the STIS instrument on the Hubble Space Telescope. 
Two examples of our fits to the SED are given in Fig. \ref{F1}, using the parameters derived by \citet{KK09}
without the benefit of the near-UV data. Nonetheless, the models fits of \citet{KK09} reproduce the near-UV flux
of all eight stars reasonably well with pure-H models, including three stars to which \citet{BLR01} assigned a He-rich composition.
The presence of hydrogen is clearly visible in the near-UV as strong Ly-$\alpha$ absorption from atomic H shortward of $\sim 4000\,\AA$ is absent in pure-He models.
This result indicates that the pure H models can be relied upon for accurate determinations of the parameters of cool WD with H-rich atmospheres.

\begin{figure}
\resizebox{\hsize}{!}{\rotatebox{270}{\includegraphics{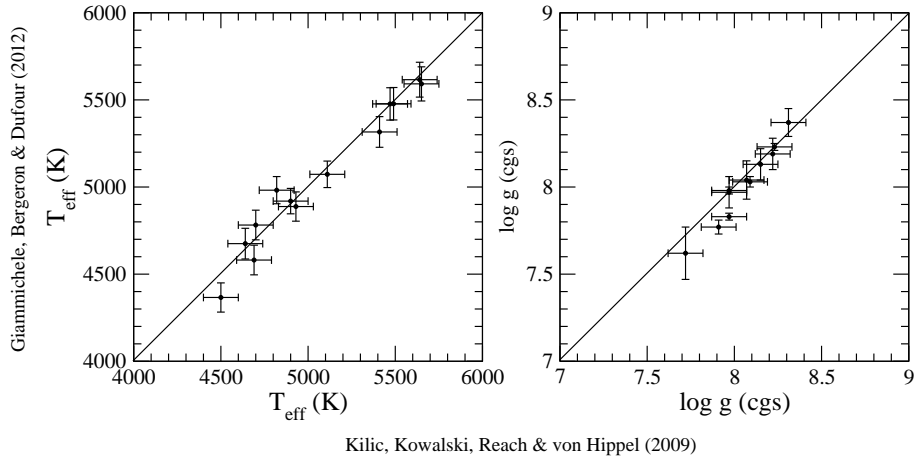}}}
  \caption{Comparison of the atmosphere parameters 
  ($T_{\rm eff}$ and surface gravity) of the stars analyzed by both \citet{KK09} and \citet{GM12}. \label{F2}}
\end{figure}

\subsection{The atmospheric composition of cool white dwarfs}

Our studies of different samples of cool white dwarfs have shown that most of the stars with $T_{\rm eff}\rm<6000\, K$ 
have hydrogen-rich atmospheres and that the amount of hydrogen increases with the age of a star (decrease of $T_{\rm eff}$, e.g. \citet{KK09,K06}).
This result was confirmed recently by \citet{GM12}, who revised the atmospheric parameters of very cool WD \citep{B97,BLR01,KB10} after including the 
Ly-$\alpha$ profiles of \citet{KS06} in their models.  Figure \ref{F2} compares $\teff$ and gravity determinations by \citet{GM12} and
\citet{KK09} for a common sample of stars. The excellent agreement between the two studies shows the reliability of state-of-the-art models for H-rich atmospheres of cool WD.

\begin{figure}
\resizebox{\hsize}{!}{\rotatebox{270}{\includegraphics{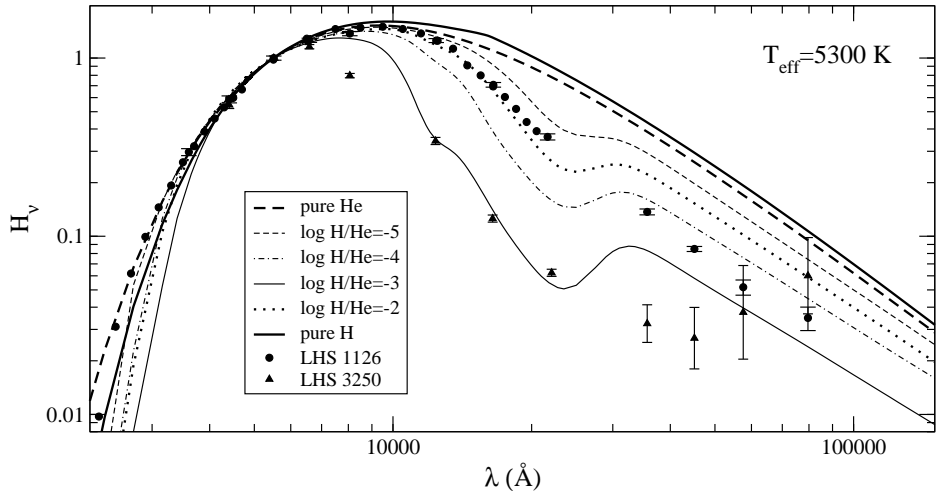}}}
  \caption{Comparison of the SED of the He-rich stars LHS 1126 and LHS 3250 with models of different $\rm H/He$ atmosphere composition, 
  $\teff=5300\,$K and $\log g=8$.  The photometric fluxes are from \citet{BLR01}, \citet{KK08} and \citet{KK09}.
  For LHS 1126 the UV flux is fromfrom  \citet{WKL02} and the near-IR flux is our new data \citep{KSL12}.
  All the spectra are normalized to the $V$ band flux. \label{F3}}
\end{figure}

\subsection{He-rich atmospheres}

Although we are able to describe well the SED of H-rich stars (Fig. \ref{F1}), He-rich atmospheres remain challenging. In Fig. \ref{F3}
we present the spectra of two He-rich stars, LHS 1126 (DQp) and LHS 3250 (DC). The presence of significant amount of hydrogen in the atmosphere of these stars is evident from the strong 
IR flux depletion caused by $\rm H_2-He$ collision-induced absorption (CIA) when comparing with the synthetic spectrum of a pure-He atmosphere. 
Although from the optical fluxes we are able to estimate that $\rm T_{\rm eff}\sim5300\,K$ for both stars, their IR spectra 
can not be well reproduced by the current models. For instance, good fits of the near-IR overestimate the IR flux beyond 3$\,\mu$m. In LHS 1126,
the near-UV spectrum suggests a different composition where He is more dominant than indicated by
the IR flux. The overall strength and frequency dependence of the IR opacity is not well reproduced by the models. 
Notably, the models predict a pronounced dip at $\sim 2.3\rm\, \mu m$ from the first overtone band of H$_2$ CIA, which is not observed.  This suggest a deficiency in the
H$_2$-He CIA under the high densities found in these He-rich atmospheres. Another possibility is
the presence of CIA from carbon species in the atmosphere of LHS 1126.
Nevertheless, until the models are able to correctly and consistently reproduce the entire SED
of WD with He-rich atmospheres, one can not draw solid conclusions regarding their atmospheric parameters, particularly the H/He ratio. More theoretical and experimental 
effort is required to correctly describe the properties of matter in these extreme atmospheres.

\section{Cool DQ/DQp white dwarfs}
\begin{figure}[t]
\resizebox{\hsize}{!}{\rotatebox{270}{\includegraphics{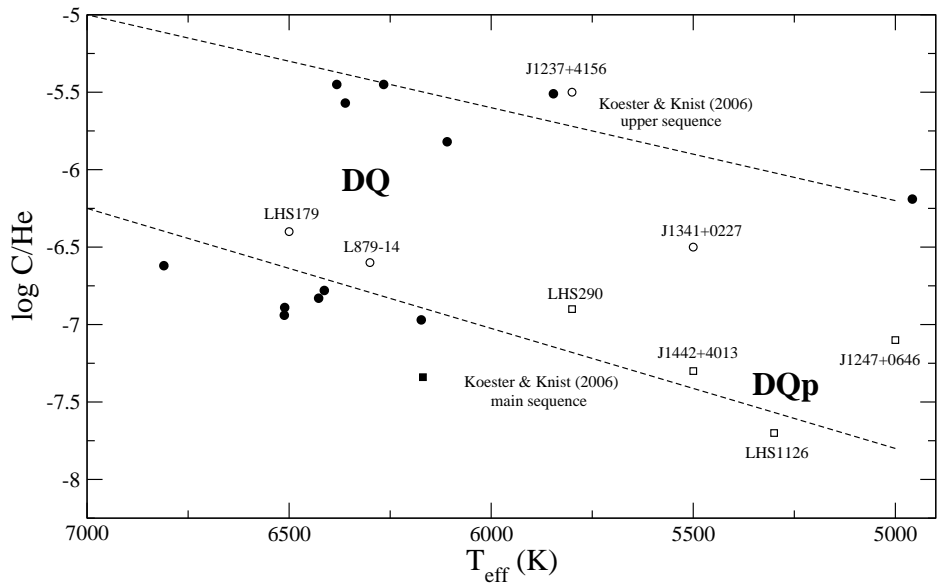}}}
  \caption{Carbon abundance in the atmospheres of cool DQ (circles) and DQp (squares) stars as a function of 
  $T_{\rm eff}$,  derived by assuming atmospheres without hydrogen. Open symbols show our results \citep{KSL12}
 and the filled symbols represent the values derived by \citet{KK06}.
 \label{F4}}
\end{figure}

Cool DQ stars have atmospheres dominated by helium with traces of carbon revealed by C$_2$ Swan bands in their spectra. The extreme character of the coolest
known examples 
manifests itself through the appearance of peculiar DQ stars (``DQp'') at $\teff \wig< 6000\,$K with distorted C$_2$-like bands \citep{SBF95,B97,K10}.
The peculiar appearance of these bands has been attributed to a new molecular species: $\rm C_2H$ \citep{SBF95,B97,BLR01}. However,
\citet{HM08} rejected this interpretation and \citet{K10} showed that the observed features are pressure-shifted $\rm C_2$ bands.
In order to investigate further the nature of DQp stars we collected and analyzed near-IR spectra of four DQ and four DQp stars \citep{KSL12}.
The spectra are featureless, which essentially rules out significant absorption from carbon-bearing molecules other than C$_2$.
We found that all DQ/DQp stars with $T_{\rm eff}\rm \wig<5500\,K$ have some hydrogen in their atmospheres, revealed by a pronounced
IR flux suppression, believed to be caused by $\rm H_2-He$ CIA \citep{KSL12}. Such a flux deficit is clearly visible in the spectrum of the DQp LHS 1126 
(Fig. \ref{F3}). However, as we already discussed in the previous section,
we are not able to fit the entire SED of this star with a single model and derive a definitive hydrogen abundance.
Nevertheless, we can determine the abundance of C in DQp stars by fitting the depth of the  distorted Swan bands in the optical with hydrogen-free models.
The C/He ratios derived for DQp stars follow the trends reported by \citet{DB05} and \citet{KK06} for DQ stars, as shown in Fig. \ref{F4}. 
\citet{BLR01}, \citet{B97} and \citet{DB05} noticed that the DQ sequence terminates abruptly at $T_{\rm eff}\rm\sim6000\,K$ and cooler stars 
with C/He$\sim 10^{-8}$ -- $10^{-7}$ are not observed, although such a carbon abundances should be
easily detectable. The results presented in figure \ref{F4} suggest that these ``missing'' DQ stars are the DQp stars. 
Note however that the C abundances derived for both DQ and DQp stars would substantially increase should hydrogen be present in their atmospheres \citep{KSL12}.
This is because even trace amounts of hydrogen increase the continuum opacity and veils the C$_2$ bands. Therefore, in mixed $\rm He/H$ atmospheres, 
more carbon is needed to produce bands of a given strength than in a hydrogen-free atmosphere. 

\section{Conclusions}
In this contribution, we gave a short overview of the current state of understanding of the atmospheres of cool white dwarfs and discussed some results of our recent observations
and application of our models.  Having analyzed the near-UV spectra of several cool white dwarfs in conjunction with their full SED, we 
conclude that hydrogen-rich atmosphere models of very cool WD are quite reliable. 
The initial claim of \citet{KS06} that the nearly all of the coolest DC white dwarfs have hydrogen-rich 
if not pure-H atmospheres is strengthened by our analysis of near-UV spectra and confirmed by an independent analysis \citep{GM12}.
Our investigation of the IR fluxes of cool DQp stars have revealed that in addition to helium and carbon, hydrogen 
is also present in their atmospheres. However, the discrepancy between the observed and modeled fluxes in the IR
and an inability to fully reproduce the entire SED of WD with mixed H/He composition dominated by He reveals a persistent deficiency in the modeled
opacities. This prevents the derivation
of reliable H abundances for the stars with mixed H/He atmospheric composition, including the DQ and DQp white dwarfs. Nevertheless, the carbon abundances 
we have derived for DQp stars support the recent claim of \citet{K10} that DQp stars are just DQ stars showing pressure-distorted C$_2$ bands 
in their optical spectra. We are addressing these difficulties by pursuing theoretical modeling of the relevant opacities in the exotic
conditions found in these atmospheres.

\acknowledgements 

Part of this work is based on observations made with the NASA/ESA Hubble Space Telescope, obtained at the Space Telescope Science Institute, and is supported by
NASA through a grant from the Space Telescope Science Institute, which is operated by the Association of Universities for Research in 
Astronomy, Inc., under NASA contract NAS 5-26555.  These observations and grant are associated with program \#HST-GO-12188. 

\bibliography{kowalski}

\end{document}